\documentstyle[12pt]{article}

\title{ Multicomponent nonisothermal nucleation. 1. Kinetic equation}

\author{V.B.Kurasov}

\date{Victor.Kurasov@pobox.spbu.ru}

\begin{document}

\maketitle

The first order phase transition ordinary occurs in the systems with many
different condensating components. During the process of nucleation, i.e.
formation of droplets, the heat of condensation is extracted which changes
the rate of nucleation.
The theoretical description has to  take into account these two features.
So, one can see that creation of the nonisothermal theory for the
multicomponent
nucleation is rather actual.

The adequate theoretical description of the nucleation stage was given
for many different situations starting from the pure isothermal nucleation
of one component (one substance). Even in this situation described by
the
so-called classical theory of nucleation there is no coincidence between
theoretical predictions and experimental results. Nevertheless, this
disagreement
can not be the reason to reject all further modifications of the theory
to grasp the effects of the heat extraction and to extend the theory for
the multicomponent case.

The first theory where the thermal effects were taken into account rather
adequately was presented by Kantrowitz  \cite{Kantrowitz}. In this
publication
embryos of given size were characterized by the unique average temperature.
Only the so-called "weak thermal effects" were considered there.
But the really essential decrease of the nucleation rate occurs under
the "strong thermal effects" considered by Kuni \cite{Kuni-prep}.

Certainly, there exists the distribution of the embryos of given size
over the temperature. Feder et al. \cite{Feder} took this phenomena into
account for the weak thermal effects.
The energy distribution of  embryos under the strong thermal effects was
investigated by Kuni and Grinin \cite{Kuni-Grinin-TMF}.

The standard way to construct  kinetic equation is to use the Fokker-Planck
approximation. Then kinetic equation can be solved in the stationary
situation by
 approach of Langer
\cite{Langer} or by approach of Kuni et al.
\cite{Kuni-mul} in the nonstationary situation.

The methods described in \cite{Langer} allow to study some different
situations. Namely  the  situation  of  the  binary  situation  was
studded
by
Lazaridis and Drossinos \cite{L-D} under the Fokker-Planck approximation.

Concerning the mentioned publication \cite{L-D} one has to note that
due to a rather big quantity of the molecular heat extracted in the act
of condensation the restriction by the Fokker-Planck approximation isn't
sufficient.
Then kinetic equation contains high derivatives and one has to use the
Chapman-Enskog procedure to solve it.

Until the current
moment one can say that the most advanced approach for the nonisothermal
nucleation is the application
of the Chapman-Enskog procedure which was suggested by Kuni, Grinin in
the case of one component nucleation \cite{Kuni-Grinin-TMF}.
In \cite{Djik} this approach was
spread to the binary nucleation. Unfortunately it was not done in the
appropriate way.
So, the subject of this publication appears rather naturally in the context
of the theoretical methods development.

Particularly, this publication is aimed to present the nonisothermal
theory of nucleation in the multicomponent mixture
and to correct the errors in \cite{Djik}.
We shall present the self consistent theory which gives the analytical
expression for the nucleation rate. Some parts of the theory where it
is quite analogous to the known results (for example, the solution of kinetic
equation in the Fokker-Planck approximation) are omitted.

In this publication we shall use all standard definitions of the classical
nucleation theory, consider the
unit volume, take all values with the energy dimension in the units of
the elementary thermal energy $k_b T$ ($k_b$ is the Boltzmann constant and
$T$ is the temperature), and all values with the heat capacity dimension
in the $k_b$ units.

\section{Substance exchange}

Here we shall derive kinetic equation from the
balance equation for the distribution $n$ over the embryos sizes. this
balance equation
can be written in the following form
$$
\frac{\partial n}{\partial t} =
V + G
$$
where
$V$ is the operator associated with the substance exchange between the
embryo and the environment (one also has to take into account  here
the energy
exchange due to the extraction of the condensation heat) and $G$ is the
operator associated with the energy exchange due to the difference of
the temperature $T$ of the embryo from the temperature $T_0$ of the
environment
(the effects of the heat extraction are already included into $V$).

The function $n$ is the number of embryos of given size in a unit volume.
Later we shall specify variables of the embryos state description.

At first we shall study kinetic equation without thermal relaxation by
a passive gas, i.e. in the form
\begin{equation} \label{.1.}
\frac{\partial n}{\partial t} =
V
\end{equation}
and then we shall study thermal relaxation, i.e. operator $G$.

Suppose that there exists a vapor mixture with  $i_0$
condensating components. All
these components can be found both in the vapor and in the embryo (in
contrary of the passive gases which can be found only in the vapor phase).
The distribution  $n(\{ \nu_i \}, t )$ of embryos  is the function of $i_0$
variables
of the number of molecules $\nu_i$      of the $i$th component
(substance) inside
the embryo and also the function of time $t$.

The process of the  absorption of a molecule of $j$th component occurs
with intensity $W^+_j \alpha_{j\ cond} $ where $W^+_j$ is the intensity of
collision
of the given embryo with an arbitrary molecule of $j$th component and
$\alpha_{j\ cond}$ is the condensation coefficient for $j$th component.
The most natural is to suppose that the length of the free motion in the
gas media strongly exceeds the linear size of the embryo (so-called free
molecular regime of the substance exchange). Then the value $W^+_j$
can be easily found by the gas kinetic theory
$$
W^+_j \sim  S  n_j
$$
where $S$ is the surface square of the embryo and $n_j$ is the molecular
number density of the $j$th component\footnote{One has also to note that
in the nearcritical region all embryos are in the quasiequilibrium with
the surrounding vapor. This quasiequilibrium also leads to the small
unhomogenities
in the   vapor  phase.  These  unhomogeneties  would  lead  to  the
diffusion flows
on the embryo. That's why one can approximately say that the
vapor around the nearcritical embryo is unperturbed and calculate the
flow by the previous formula.
}.
Due to the small relative size of the nearcritical region\footnote{One
can use the standard classical nucleation theory for this estimate.}
we can see that $W^+_j$ remains
practically constant in the nearcritical region.

Any reliable information about the condensation coefficient $\alpha_{j\ cond}$
is absent (in the literature one can found rather different estimates).
But we believe that $\alpha_{j\ cond}$ is rather smooth function of the
embryo
state  and can be regarded in the nearcritical region as some
constant value\footnote{This follows also from the model where
$\alpha_{j\ cond}$
appears as the probability to overcome the
 energy barrier near the surface of the liquid phase.}.

The process of absorption of the molecule leads
to the variation of the number of the molecules inside the embryo
$$
\{\nu_i\} \rightarrow \{\nu_{i\ne j}, \nu_j + 1 \}
$$
and also
to the extraction of the
condensation heat $\beta_j$ (which is measured in the natural thermal
units) which is going to increase the temperature of the embryo $T$ by
the value
$$
T \rightarrow T + \frac{\beta_j}{\sum_i c_i \nu_i}
$$
where $c_i$ are the heat capacities per one molecule in the liquid phase
(taken in units of $k_b$), the sum is taken over all components.

The value $W_j^+$ depends only on the state of a vapor-gas mixture.
Contrary to $W^+_j$ the intensity of the ejection of the molecule of the
$j$th component $W_j^-$ strongly depends on the temperature of the embryo.
The embryo has to be characterized by the temperature $T$ of the
embryo\footnote{We
suppose that the temperature relaxation inside occurs very rapidly. It
can be justified by estimates analogous to \cite{Kuni-Grinin-TMF}. } or
some function of the temperature.
Instead of $T$ one can introduce the value of additional energy $E$ according
to
$$
E=(\frac{T}{T_0} -1)\sum c_i \nu_i
$$
where
$T_0$ is the temperature of the media, $c_i$ are molecular specific
heats\footnote{They
are close to those defined under the constant pressure.}
expressed in units of $k_b$. The evident advantage of $E$ is that the
equilibrium value coincides with the zero point. Now we shall normalize
$E$ in order to have no coefficient in
the square form of the equilibrium distribution\footnote{We shall omit
the normalizing factor $\pi^{-1/2}$ of the equilibrium distribution. It
can be easily reconstructed anywhere.} $n^e$ along additive energy
 near $E=0$:
$$
n^e \sim \exp(-\mu^2)
$$
To find $\mu$ we shall start with Clapeyron-Clausius formula for the
molecular number density
$n_{\infty \ j }$ of the saturated vapor under a plane surface of liquid
$$
n_{\infty \ j} (T) = n_{\infty \ j } (T_0) \exp(\beta_j \frac{T-T_0}{T_0}
)
$$
As far as\footnote{/here we suppose that the surface tension doesn't depend
on the embryos temperature. Details of this approximation can be found
in \cite{Kuni-Grinin-Shekin-PhysA}. } for the planar surface
$$
W^-_j (n_{\infty \ j }) = W^+_j ( n_{\infty \ j } ) \sim n_{\infty \ j }
$$
one can come to\footnote{More carefully one has to go at first from the
critical embryo to the plane surface, then use the mentioned relation
for the plane surface and finally return to the critical embryo back.}
$$
\frac{W^-_j(T)}{W^-_j(T_0)} = \frac{n_{\infty \ j }(T)}{n_{\infty \ j }(T_0)}
= \exp(\beta_j \frac{T-T_0}{T_0})
$$

One the other hand
$$
W^-_j (E) n^e(E) = W^+_j n^e (E-\beta_j) n^e(E-\beta_j)
$$
and
$$
W^-_j (E=0) n^e(E=0) = W^+_j n^e (-\beta_j) n^e(-\beta_j)
$$
which leads to
$$
W^-_j (E) = \frac{n^e(E-\beta_j)}{n^e(E)} W^-(E=0)
\frac{n^e(E=0)}{n^e(-\beta_j)}
= W^-(E=0) \exp(E \beta_j \frac{\partial^2 F}{\partial E^2})
$$
where
$F$ is the free energy of the embryos formation and it is taken into account
that the equilibrium distribution $n^e \sim \exp(-F)$.

Both approaches will coincide when
$$
\frac{\partial^2 F}{\partial E^2} = (\sum_i c_i \nu_i)^{-1}
$$

Then\footnote{More carefully it can be done in terms of the finite
differences.}
$$
\mu = \frac{E}{(2\sum_i c_i \nu_i)^{1/2}}
$$

One can write kinetic equation (\ref{.1.}) in variables $\{ \nu_i \} , \mu$ in
the
following form

\begin{eqnarray}
\frac{\partial n(\{\nu_i \}, \mu, t) }{\partial t} =
\sum_j W^+_j n(\{ \nu_{i \ne j}, \nu_j -1 \} , \mu - \tau_j)
- \sum_j W^-_j n(\{ \nu_{i \ne j} , \nu_j \} , \mu )
\nonumber
\\
\nonumber
-\sum_j W^+_j n(\{ \nu_{i \ne j}, \nu_j \} , \mu )
+\sum_j W^-_j n(\{ \nu_{i \ne j}, \nu_j +1 \}, \mu + \tau_j)
\end{eqnarray}
where
$$
\tau_j = \frac{\beta_j}{(\sum_i c_i \nu_i)^{1/2}}
$$

One can present the following split of the last equation
$$
\frac{\partial n(\{\nu_i \}, \mu, t) }{\partial t} =
\sum_j [ J_j ( \{ \nu_{i\ne j},  \nu_j-1 \},  \mu - \tau_j) -
J_j(\{ \nu_{i\ne j}, \nu_j \} , \mu) ]
$$
where the flow $J_j$ is defined by
$$
J_j(\{ \nu_i \},  \mu)                      =
 W^+_j n(\{ \nu_{i \ne j} , \nu_j \}  , \mu )
- W^-_j n(\{ \nu_{i \ne j}, \nu_j+1 \} , \mu+\tau_j )
$$
Now we have to substitute the finite differences by derivatives.

One has to mention that the elementary steps $1$ along
$\nu_i$ are small in comparison with the characteristic scale corresponding
to the essential variation of exponent of the free energy. This allows to
substitute
the finite difference along $\nu_i$ only by the first derivative\footnote{This
produces certain restrictions which will limit later the Chapman-Enskog
expansion}.

An elementary step $\tau_j$ along $\mu$ corresponds to essential
violation of exponent of the free energy. So, one has to substitute the
finite difference by the whole Tailor seria.  As the result we have
\begin{eqnarray}
\frac{\partial n(\{\nu_i \}, \mu, t) }{\partial t} =
\sum_j \sum_{l=1}^{\infty} \frac{(-\tau_j)^l}{l!} \frac{\partial^l}{\partial
\mu^l} J_j(\{\nu_{i\ne j }, \nu_j-1 \}, \mu)
\nonumber
\\
\nonumber
- \sum_j \frac{\partial}{\partial \nu_j }  J_j (\{\nu_{i\ne j }, \nu_j \},
\mu)
\end{eqnarray}

We have to note that the possibility to substitute
$ J_j (\{\nu_{i\ne j }, \nu_j-1 \}, \mu) -  J_j (\{\nu_{i\ne j }, \nu_j
\}, \mu)$
by
$- \frac{\partial}{\partial \nu_j }  J_j (\{\nu_{i\ne j }, \nu_j \}, \mu)$
can be made when we consider the situation near the quasistationary one.
Then in one dimensional projection on $\nu_j$ we shall get the small value
for
$ J_j ( \{ \nu_{i\ne j }, \nu_j-1 \}, \mu) -  J_j(\{\nu_{i\ne j }, \nu_j
\}, \mu)$
which allows to substitute it only by the first derivative.

The flow $J_j $ can be  expressed with the help of a function
$$
f(\{\nu_i \}, \mu) = \frac{
n(\{\nu_i \}, \mu)         }
                      {n^e (\{\nu_i \}, \mu)}
$$
as
$$
 J_j (\{\nu_{i\ne j }, \nu_j \}, \mu) =
W_j^+  n^e (\{\nu_i \}, \mu)
[ f(\{\nu_{i\ne j }, \nu_j \}, \mu)  -
f ( \{ \nu_{i\ne j }, \nu_j+1 \}, \mu + \tau_j) ]
$$
The analogous substitution of the finite differences by the Taylor seria
gives
\begin{eqnarray}
 J_j (\{\nu_{i\ne j }, \nu_j \}, \mu) =
W_j^+  n^e (\{\nu_i \}, \mu, t)
\nonumber
\\
\nonumber
[ - \frac{\partial}{\partial \nu_j }  f(\{\nu_{i\ne j }, \nu_j+1 \}, \mu)
- \sum_{m=1}^{\infty} \frac{\tau_j^m}{m!} \frac{\partial^m}{\partial \mu^m}
 f(\{\nu_{i\ne j }, \nu_j+1 \}, \mu)  ]
\end{eqnarray}

To be close to the standard form of one component nonisothermal theory
\cite{Kuni-Grinin-TMF} we shall use instead of function $f$ the following
function
$$
P(\{\nu_{i\ne j }, \nu_j \}, \mu)
 = \frac{ n(\{\nu_{i\ne j }, \nu_j \}, \mu) }{\exp( - \mu^2)}
$$

The free energy $F$ of the embryos formation can be split as
$$
F(\{\nu\}, \mu) = F (\{\nu\}, \mu=0) +  \mu^2
$$
which gives
$$
f(\{\nu\}, \mu) =   P(\{\nu\}, \mu) \exp(F(\{\nu\}, \mu=0 ) )
$$

One can present expression for $J_j$ in terms of function $P$ as
\begin{eqnarray}
 J_j (\{\nu_{i\ne j }, \nu_j \}, \mu) =
W_j^+  n^e (\{\nu_i \}, \mu)
[ - \frac{\partial}{\partial \nu_j }
 \exp(F(\{\nu\}, \mu=0 )  P(\{\nu_{i\ne j }, \nu_j \}, \mu)
-
\nonumber
\\
\nonumber
\exp( F ( \{ \nu_{i \ i\ne j}, \nu_j+1 \}, \mu=0 )
 \sum_{m=1}^{\infty} \frac{\tau_j^m}{m!} \frac{\partial^m}{\partial \mu^m}
 P(\{\nu_{i\ne j }, \nu_j+1 \}, \mu)  ]
\end{eqnarray}
or
\begin{eqnarray}
 J_j (\{\nu_{i\ne j }, \nu_j \}, \mu) =
W_j^+  n^e (\{\nu_i \}, \mu)
[ - \frac{\partial}{\partial \nu_j }
 \exp(F(\{\nu\}, \mu=0 )  P(\{\nu_{i\ne j }, \nu_j \}, \mu)
-
\nonumber
\\
\nonumber
(1+\frac{\partial F}{\partial \nu_j} +
\frac{1}{2} \frac{\partial^2 F}{\partial \nu_j^2}
+
\frac{1}{2} (\frac{\partial F}{\partial \nu_j})^2
)
\exp( F ( \{ \nu_{i \ i\ne j}, \nu_j \}, \mu=0 )
 \sum_{m=1}^{\infty} \frac{\tau_j^m}{m!} \frac{\partial^m}{\partial \mu^m}
 P(\{\nu_{i\ne j }, \nu_j+1 \}, \mu)  ]
\end{eqnarray}

Having introduced an operator
$$
L_j = - W^+_j [ \frac{\partial F}{\partial \nu_j} + \frac{\partial}{\partial
\nu_j} ]
$$
one can present the last expression for $J_j$ as
\begin{eqnarray}
\nonumber
 J_j (\{\nu_{i\ne j }, \nu_j \}, \mu) =
 \exp(-\mu^2)  L_j P(\{\nu_{i\ne j }, \nu_j \}, \mu)
-
\\
\nonumber
(1+\frac{\partial F}{\partial \nu_j} +
\frac{1}{2} \frac{\partial^2 F}{\partial \nu_j^2}
+
\frac{1}{2} (\frac{\partial F}{\partial \nu_j})^2
)
\exp(-\mu^2)       W_j^+
 \sum_{m=1}^{\infty} \frac{\tau_j^m}{m!}
\frac{\partial^m}{\partial \mu^m}
 P(\{\nu_{i\ne j }, \nu_j+1 \}, \mu)  ]
\end{eqnarray}

According to the smooth dependence along $\nu_j$  one can substitute the
argument $\nu_j \pm 1$ of function\footnote{Later it will be seen that
$\partial / \partial \nu_j$ produces some small parameter as far as the
action of $\partial F / \partial \nu_j$. We write this equation in the
first two orders of this parameter. This corresponds to the order essential
in the isothermal version of the theory. } $P$  by $\nu_j$.

Then
as far as
$$
(1+\frac{\partial F}{\partial \nu_j} +
\frac{1}{2} \frac{\partial^2 F}{\partial \nu_j^2}
+
\frac{1}{2} (\frac{\partial F}{\partial \nu_j})^2
)|_{\nu \rightarrow \nu+1} =
(1+\frac{\partial F}{\partial \nu_j} -
\frac{1}{2} \frac{\partial^2 F}{\partial \nu_j^2}
+
\frac{1}{2} (\frac{\partial F}{\partial \nu_j})^2
)
$$
then kinetic equation can be presented as
\begin{eqnarray}
\exp(-\mu^2) \frac{\partial  P(\{\nu_i \},  \mu) }{\partial t} =
\nonumber
\\
\sum_j \sum_{l=1}^{\infty} \frac{(-\tau_j)^l}{l!}
\frac{\partial^l}{\partial \mu^l}
\nonumber
\\
\nonumber
[ \exp(-\mu^2) L_j
(1 - \frac{\partial}{\partial \nu_j} )
P(\{\nu_i \},  \mu)  -
\\
\nonumber
(1+\frac{\partial F}{\partial \nu_j} -
\frac{1}{2} \frac{\partial^2 F}{\partial \nu_j^2}
+
\frac{1}{2} (\frac{\partial F}{\partial \nu_j})^2
)
\exp(-\mu^2) W^+_j  \sum_{m=1}^{\infty} \frac{\tau_j^m}{m!}
\frac{\partial^m}{\partial \mu^m} P(\{ \nu_i , \mu \}) ]
-
\\
\nonumber
\sum_j
\frac{\partial}{\partial \nu_j }
[ \exp(-\mu^2) L_j  P(\{\nu_i \},  \mu)  -
(1+\frac{\partial F}{\partial \nu_j})
(1+\frac{\partial  }{\partial \nu_j})
\exp(-\mu^2) W^+_j  \sum_{m=1}^{\infty} \frac{\tau_j^m}{m!}
\frac{\partial^m}{\partial \mu^m} P(\{ \nu_i , \mu \}) ]
\end{eqnarray}

The action of $\partial / \partial \mu$ on $\exp(-\mu^2) \psi$ where
$\psi $ is the arbitrary function is obviously given by
$$
\frac{\partial^l}{\partial \mu^l}
\exp(-\mu^2) \psi =
exp(-\mu^2) (
\frac{\partial^l}{\partial \mu^l} -
2 \mu)^l
\psi
$$
Certainly one can not take $2 \mu$  away from
$
\frac{\partial^l}{\partial \mu^l}$
and has to consider $ (
\frac{\partial^l}{\partial \mu^l} -
2 \mu)^l $ as sequential action of  operators in brackets.
This turn  kinetic equation to
\begin{eqnarray}
 \frac{\partial  P(\{\nu_i \},  \mu) }{\partial t} =
\sum_j \sum_{l=1}^{\infty} \frac{(-\tau_j)^l}{l!}
(\frac{\partial}{\partial \mu} - 2\mu)^l
\nonumber
\\
\nonumber
[ L_j   (1- \frac{\partial}{\partial \nu_j } ) P(\{\nu_i \},  \mu)  -
 W^+_j
(1+\frac{\partial F}{\partial \nu_j} -
\frac{1}{2} \frac{\partial^2 F}{\partial \nu_j^2}
+
\frac{1}{2} (\frac{\partial F}{\partial \nu_j})^2
)
\sum_{m=1}^{\infty} \frac{\tau_j^m}{m!}
\frac{\partial^m}{\partial \mu^m} P(\{ \nu_i , \mu \}) ]
-
\\
\nonumber
\sum_j
\frac{\partial}{\partial \nu_j }
[  L_j  P(\{\nu_i \},  \mu)  -
 W^+_j
(1+ \frac{\partial F}{\partial \nu_j} +
\frac{\partial}{\partial \nu_j} )
 \sum_{m=1}^{\infty} \frac{\tau_j^m}{m!}
\frac{\partial^m}{\partial \mu^m} P(\{ \nu_i , \mu \}) ]
\end{eqnarray}

The last equation present the final form of kinetic equation.

We  decompose  the finite differences along $\nu_j$  until the  second
derivatives (or the second order of the  small parameter) because  in
the classical theory of isothermal one-component nucleation two derivatives
have to be taken into account (An account
of the first derivative
couldn't lead to the suitable rate of nucleation).

One can easily note that operators
$$
S_{1j} = 1-\frac{\partial}{\partial \nu_j}
$$
$$
S_{2j} =
1+\frac{\partial F}{\partial \nu_j} -
\frac{1}{2} \frac{\partial^2 F}{\partial \nu_j^2}
+
\frac{1}{2} (\frac{\partial F}{\partial \nu_j})^2
$$
$$
S_{3j} =
1+
\frac{\partial F}{\partial \nu_j}
+
\frac{\partial }{\partial \nu_j}
$$
are absent in \cite{Djik}. Really, in the first two steps of the
Chapman-Enskog
procedure described later these terms will not be essential. But it occurs
only in frames of the Chapman-Enskog procedure and can not be seen directly
from this equation. We shall call $S_{1j}$, $S_{2j}$, $S_{3j}$ as the
shift operators.

Now we shall
turn to the thermal relaxation by the passive gas in order to include
it in  the presented equation.

\section{Thermal relaxation}

The physical reason to consider the interaction of the embryo with the
passive gas is rather simple. Really, due to  heat extraction the temperature
of the embryo is higher than the temperature of environment and the embryo
heats the molecules of the passive gas. Certainly, the temperature of
the embryo falls which reduces an ejection rate. This has to be taken
into account and the consideration of the interaction with the passive
gas is important.

In the previous consideration the condensation of the molecule can be
described by some fixed values of the the condensation heat $\beta_j$.
An obvious restriction only by the regular term  in the presence
of the big quantities of the passive gas will lead to the thin spectrum
in the energy scale  of the $\delta -$function  form. Certainly, this doesn't
coincide
with the equilibrium distribution. Thus, one has to use at least the
Fokker-Planck
approximation.
The physical reason is rather obvious -  molecules of a passive have the
different velocities and equilibrium distribution in energies. This has
to be taken into account and leads at least to the Fokker-Planck approximation.
As far as the variation of the energy  in  the  elementary  act  of
interaction
is small in comparison with the characteristic scale of the variation
of the equilibrium distribution one can restrict this description by the
Fokker-Planck approximation.

In Fokker-Planck approximation the kinetic equation can be written as
$$
\frac{\partial n}{\partial t} =
B
\frac{\partial}{\partial \mu} n^e
\frac{\partial}{\partial \mu} f
$$
where $B$ is kinetic coefficient. It can be determined by consideration
of the limit situation where the last equation transforms into an equation
only with the regular term
$$
\frac{\partial n}{\partial t} =
B
\frac{\partial}{\partial \mu}
[ 2 \mu + \frac{\partial}{\partial \mu}  ] n
\rightarrow
B
\frac{\partial}{\partial \mu}
2 \mu   n
$$

This form form has to be reproduced by the standard analysis. We begin
with the balance equation
$$
\frac{\partial n}{\partial t} =
W^+ n(\mu+ \delta \mu) - W^+ n(\mu)
\rightarrow W^+ \delta \mu \frac{\partial n}{\partial \mu}
$$
where
$W^+$ is the rate of collisions of the given embryo with the molecules
of the passive gas, the regular variation $\delta \mu$ is given by
$$
\delta \mu = \frac{c_g}{\sum c_j \nu_j} \mu
$$
and $c_g$ is the molecule heat capacity of the passive gas.
This leads to
$$
B = W^+ \frac{c_g}{2 \sum c_j \nu_j}
$$

Now the kinetic equation can be written in the following form
$$
\frac{\partial n}{\partial t} =
 W^+ \frac{c_g}{2 \sum c_j \nu_j}
\frac{\partial}{\partial \mu} n^e
\frac{\partial}{\partial \mu} f
$$
One has to put the coefficient of thermal accommodation $\alpha_{acc}$ to
$W^+$ in order to take into account that thermal accommodation occurs with
some probability.

The generalization of the previous equation on the mixture of passive
gases leads to
$$
\frac{\partial n}{\partial t} =
\sum_{i'}
 W^+_{i'} \alpha_{acc\ i'} \frac{c_{g\ i'}} {2 \sum c_j \nu_j}
\frac{\partial}{\partial \mu} n^e
\frac{\partial}{\partial \mu} f
$$
where indexes with prime denote different passive gases.

One has to take into account that the condensating substances also take
part in the cooling. With the probability $(1-\alpha_c)
\alpha_{acc}$ the act of cooling takes place. The molecules
accumulated
by embryos also have to be taken into account. As the result one can get
\begin{eqnarray}
\frac{\partial n}{\partial t} =
\sum_{i'}
 W^+_{i'} \alpha_{acc\ i'} \frac{c_{g\ i'}} {2 \sum c_j \nu_j}
\frac{\partial}{\partial \mu} n^e
\frac{\partial}{\partial \mu} f
+
\nonumber
\\
\nonumber
\sum_{i}
 W^+_{i} (1-\alpha_{c\ i}) \alpha_{acc\ i} \frac{c_{i}} {2 \sum c_j \nu_j}
\frac{\partial}{\partial \mu} n^e
\frac{\partial}{\partial \mu} f
+
\\
\nonumber
\sum_{i}
 W^+_{i} \alpha_{c\ i} \frac{c_{i}} {2 \sum c_j \nu_j}
\frac{\partial}{\partial \mu} n^e
\frac{\partial}{\partial \mu} f
\end{eqnarray}

It is quite obvious that now to get the general kinetic equation
we have to add the part associated with the
condensating substance. So, it is necessary to formulate the part under
consideration in terms of function $P$. Here one has to fulfill the same
actions and get
\begin{eqnarray}
\nonumber
  \frac{\partial P}{\partial t} =
\sum_{i'}
 W^+_{i'} \alpha_{acc\ i'} \frac{c_{g\ i'}} {2 \sum c_j \nu_j}
(\frac{\partial}{\partial \mu}-2\mu)
\frac{\partial}{\partial \mu} P
+
\\
\nonumber
\sum_{i}
 W^+_{i} (1-\alpha_{c\  i}) \alpha_{acc\ i} \frac{c_{i}} {2 \sum c_j \nu_j}
(\frac{\partial}{\partial \mu}-2\mu)
\frac{\partial}{\partial \mu} P
+
\\
\nonumber
\sum_{i}
 W^+_{i} \alpha_{c\ i} \frac{c_{i}} {2 \sum c_j \nu_j}
(\frac{\partial}{\partial \mu}-2\mu)
\frac{\partial}{\partial \mu} P
\end{eqnarray}

 As the final result for the general kinetic equation
one can get
\begin{eqnarray}
\nonumber
 \frac{\partial  P(\{\nu_i \},  \mu) }{\partial t} =
\nonumber
\\
\sum_j \sum_{l=1}^{\infty} \frac{(-\tau_j)^l}{l!}
(\frac{\partial}{\partial \mu} - 2\mu)^l
[ L_j  S_{1j} P(\{\nu_i \},  \mu)  -
 S_{2j} W^+_j  \sum_{m=1}^{\infty} \frac{\tau_j^m}{m!}
\frac{\partial^m}{\partial \mu^m} P(\{ \nu_i \}, \mu) ]
-
\\ \nonumber
\sum_j
\frac{\partial}{\partial \nu_j }
[  L_j  P(\{\nu_i \},  \mu)  -
 S_{3j} W^+_j  \sum_{m=1}^{\infty} \frac{\tau_j^m}{m!}
\frac{\partial^m}{\partial \mu^m} P(\{ \nu_i \}, \mu) ]
+
\\ \nonumber
\sum_{j'}
 W^+_{j'} \alpha_{acc\ j'} \frac{c_{g\ j'}} {2 \sum_j c_j \nu_j}
(\frac{\partial}{\partial \mu}-2\mu)
\frac{\partial}{\partial \mu} P(\{ \nu_i \}, \mu)
+
\\ \nonumber
\sum_{i'}
 W^+_{i'} (1-\alpha_{c\ i'}) \alpha_{acc\ i'} \frac{c_{i'}} {2 \sum_j c_j
\nu_j}
(\frac{\partial}{\partial \mu}-2\mu)
\frac{\partial}{\partial \mu} P(\{ \nu_i \}, \mu)
+
\\ \nonumber
\sum_{i'}
 W^+_{i'} \alpha_{c\ i'} \frac{c_{i'}} {2 \sum_j c_j \nu_j}
(\frac{\partial}{\partial \mu}-2\mu)
\frac{\partial}{\partial \mu}
 P(\{ \nu_i \}, \mu)
\end{eqnarray}
Here indexes $i'$ and $j$ mark the different components of the condensating
substances and index $j'$  marks the different components of the passive
substances.

One can easily note that the number of components of the condensating
mixture doesn't act on the  properties of  passive gases.
This lies in contradiction with results presented in \cite{Djik}
where the action of the passive gas is referred to the action of every
component of condensating substance and then the direct summation over the
condensation components is carried out. So, according to \cite{Djik} one
can speak about the separate cooling of different components (in kinetic
sense, the droplet is cooling as a whole object).
Here we speak about the common cooling of different components (in kinetic
sense). The physical essence is another here.

\section{Estimates of operators}

Now we shall present the method to solve the last equation. The general
scheme is well known. At first the  extraction of the main operator with
the well known eigenfunctions has to be  presented.
This main operator has to ensure the relaxation to the stationary state
which allows to  consider of the relaxation period.
Such a structure   allows later to apply the Chapman-Enskog procedure.

An attempt to investigate the situation of the binary nonisothermal nucleation
was made in \cite{Djik} but the initial kinetic equation was wrong.
 Contrary to  \cite{Djik} we shall use the
correct kinetic equation. Here this equation is
already generalized for the
multicomponent case.

Besides the new object of investigation the approach presented here has
also some new principal features.

    One has to note the specific feature of the relaxation stage description
fulfilled in the mentioned papers.
When the main operator has only formal priority
then a standard  consideration of the relaxation stage requires the small
value of specific parameter.
 This leads to the serious restriction of  the approach
used in the mentioned papers.
In the situations of intensive droplets formation this parameter isn't
too small and the relaxation  doesn't take
place. Then the initial condition for the Chapman-Enskog procedure is
violated. This doesn't allow to apply this procedure.

We shall use another split on the r.h.s. of kinetic equation into the
main operator and the additive one. As the result we come to some more
complicate procedure with two sets of the main and additive operators.
But still in such situation it will be possible to generalize the
Chapman-Enskog
procedure and come to the final formulas. It will be possible to get the
relaxation to the stationary state without restriction used in \cite{Djik}.
One has to mention that the cited papers couldn't overcome the main
nontrivial feature of the nonisothermal condensation - the main operator
extracted in these papers has only formal priority based on the presence
of factorials in denominators in the Taylor's expansion terms.
So, one has to fulfill at least many  steps in the Chapman-Enskog procedure.
Here we shall present the way how to take into account the tails of these
series and to come to the compact final results.

Now it is worth mentioning the inclusion of the present analysis into
the general scheme.

One can note some specific features of the thermal effects in comparison
with the general situation of the non Fokker-Planck evolution considered
in \cite{book}. Namely these features allow us to go further in comparison
with \cite{book} and to get the compact final formulas.

These features are
the following ones:
\begin{itemize}
\item
The temperature of the embryo can have an arbitrary value.
\item
Non-Fokker-Planck evolution occurs along the temperature of the embryo.
\item
Non-Fokker-Planck evolution  occurs under the constant value of $\beta_j$.
\end{itemize}

The third feature is rather important. Really, as far as we have the
Clapeyron-Clausius
relation we can reconstruct $W^-(T)$. Then on the base of $W^+$
(it is given by the simple gas kinetics formula) and $W^-$ one can get
the equilibrium distribution. The knowledge of the equilibrium distribution
gives on the base of the Boltzmann formula the form of the free energy $F$
of
the embryo formation  (the constant shift appeared from the normalizing
factor of the  equilibrium  distribution  isn't  important).  These
constructions
result in a rather simple form of the free energy. In the arbitrary situation
(see \cite{book})
the form of the free energy can be more complicated and this causes the
additional difficulties.

The mentioned simple form of the free
energy corresponds to the simplicity of transition from the function $n$ to
the function $P$ defined by
$$
n = \exp(-\mu^2) P
$$
Certainly $\exp(-\mu^2)$ represents here the equilibrium distribution
and $\mu^2$ appears due to the square character of the free energy.
This leads to the $P$ relaxation to a constant.
To conserve  such
relaxation
in the general more complicated situation one has to choose
in the argument of exponent instead of $\mu^2$
another more complicated function
which reflects the more complicated behavior of the free energy.
As the result the eigenfunctions of the "main" operator will be unknown.

In the general situation
 instead of $\mu$ in the combinations
  $- 2 \mu + \partial /
\partial \mu$ in the kinetic equation appear high powers of $\mu$. This blocks
the
presented approach to get  solution.

We are going to act in frames of the macroscopic description of the free
embryo. This leads to the big parameters
$$
\nu_{i\ c} \gg 1
$$
for all
components which are marked by index $i$. Index $c$ corresponds to the
critical embryo.

The last inequality allows as it is shown in \cite{book} to state that
\begin{itemize}
             \item
The Fokker-Planck approximation is valid to describe the evolution along
$\nu_i$
\item
The square approximation for the free energy
along $\nu_i$ in the near critical region
is valid.
\end{itemize}

In \cite{book} all specific situations appear only when the derivative
of the free energy along the concentration of the solution inside the
embryo provides another big parameter. Certainly the essential values
of this derivative are rather ordinary in the nature  but  one  can
not consider
them as the big parameter going to infinity.

Really the derivative along $\nu_i$ has a big value in comparison with
the derivative along the steepens descent line \cite{book}.
This appears as the base for the hierarchy in the nearcritical region
\cite{book}. But the value of this big parameter isn't sufficient
to compensate
the influence of the big parameter $\nu_i$ (as far as all $\nu_i$ has
one and the same power we shall drop the index $i$ in the estimates).
Namely,
the halfwidth $
\Delta \nu$  along $\nu$
 has the order $\nu_c^{1/2}$, the halfwidth along the steepens
descent line has the order $\nu_c^{2/3}$.
But as far as $\Delta \nu$ is greater than $1$ (it isn't so great as in the
one component theory but it is still great) we can see that the differential
form of the kinetic equation is valid.

The last result can be directly seen from the explicit expression for
the free energy of the embryos formation as the function of $\{ \nu_i
\}$. In the capillary approximation this expression can be written as
$$
F \sim - \sum_j b_j \nu_j + a (\sum_j v_j \nu_j)^{2/3}
$$
Here $v_j$ are the molecule volumes in the liquid phase, $b_j$ are the
excesses
of the chemical potentials, $a$ is the renormalized surface tension.
The surface of tension is put as to contain precisely the volume of the
embryo.
All $v_j$ are supposed to have one and the same order, all $\nu_c$ are
also supposed to have one and the same order.

One can easy note that the halfwidth $\Delta  \nu \sim \nu^{1/2}$
in the multicomponent
theory differs from the same value in the one component theory
$\Delta \nu \sim \nu^{2/3}$. The reason is the interaction between components.
This phenomena doesn't lie in contradiction with the general theory because
the steepens descent line doesn't coincide with any $\nu_i$ and the halfwidth
along the steepens descent line coincides with the halfwidth along $\nu$
in the one component theory. But it shows that the direct differentiation
of $F$ along $\nu_i$ without the influence of the other components taken
into account cannot lead  to the really small parameter. An account
of the mentioned interaction is rather
difficult and it is more convenient to go to variables $\xi_i =
\nu_i / \sum_j \nu_j$ and $\kappa = a^{3/2} \sum_j \nu_j v_j$.
Then the direction along $\kappa$ coincides with the steepens descent
line and due to the Gibbs-Duhem equation an account of the interaction
is attained automatically.

In the set $\{ \xi_i \}, \kappa $ the form of the free energy is given
by
$$
F \sim B \kappa - \kappa^{2/3}
$$
where $B$ is some function of $\{ \xi_i \}$.
The characteristic scale of $\kappa$ can be put as to
coincide with the scale of $\nu_i$.

To justify the validity of the square approximation one has to get the
second and the third derivatives  of the free energy.
It is more convenient to use the last form of $F$. Then
$$
\frac{\partial^2 F}{\partial \xi_i \partial \xi_j }  \sim
\frac{ d^2 B}{d\xi_i d\xi_j} \kappa
$$
and the halfwidth along $\xi_i $ is given by
$$
\Delta \xi_i \sim \kappa^{-1/2} (B''(\{ \xi_j \} ))^{-1/2}
$$

As far as
$$
\frac{\partial^3  F}{\partial \xi_i \xi_j \xi_k } \sim
\frac{d^3 B}{d\xi_i d \xi_j d \xi_k} \kappa
$$
the action of the third term in the Taylor seria is given by
$$
\frac{1}{3!} \frac{\partial^3 F}{\partial \xi_i^3}
(\Delta \xi_i)^3
\sim
\frac{B'''(\{\xi_j\} ) }{(B''(\{\xi_j\}))^{3/2}}
\kappa^{-1/2}
$$
As far as the function $B$ and it's derivatives don't contain any big
parameter one can easy see that the action of the third term is small.
That's why the square
approximation for the behavior of $F$ along $\xi_i$ is valid.
The behavior of $F$ along $\kappa$ is similar to the one-component case.
The square approximation along $\kappa$ is, thus, valid.
As the result the square approximation for $F$ in the nearcritical region
can be used.

These results explain why the Fokker-Planck approximation is adopted for
the description of the evolution along $\nu_i$. As for the evolution along
 temperature the Fokker-Planck approximation isn't sufficient. The reason
is the existence of another big parameter. This parameter is $\beta_i$.

Let's explain why\footnote{The index will be omitted.}
$\beta$ can be regarded as the big parameter of the
theory. When the temperature decreases from the value of the second order
phase transition the value of $v_v$ of the molecule volume in the vapor
phase
grows and the value $v_l$ falls. So, far from the point of the second
order phase transition\footnote{Namely in this situation some actual
assumptions
of the classical theory of nucleation are valid, for example the
uncompressibility
of liquid phase.} one can come to
$$
v_v \gg v_l
$$

The heat extracted in the phase transition can be presented as the difference
of entropies in two phases multiplied by temperature.
Then one can use the standard representation
of the entropy as the logarithm of the states number.
The number of states\footnote{One can use quasiclassical approach.}
 can be very approximately
estimated as the volume occupied by the system\footnote{We can use very
approximately the
model of ideal gas.}. Then the
last strong inequality leads to the big value of $\beta$.

In reality one can not go very far from the second order phase transition
temperature because the new phase transition (crystallization) occurs.
Nevertheless one has to say that $\beta$ is the big parameter of the theory.

Now we come to the direct solution of the kinetic equation. We can rearrange
it in the following form
\begin{equation}
 \frac{\partial  P(\{\nu_i \},  \mu) }{\partial t} =
D_1 + D_2 + D_3 + D_4
\end{equation}

$$
D_4 =
-
\\
\sum_j
\frac{\partial}{\partial \nu_j }
  L_j  P(\{\nu_i \},  \mu)
$$

$$
D_2 = - \sum_j \sum_{l=1}^{\infty} \frac{(-\tau_j)^l}{l!}
(\frac{\partial}{\partial \mu} - 2\mu)^l
 W^+_j  S_{2j} \sum_{m=1\  (m \ne l)}^{\infty} \frac{\tau_j^m}{m!}
\frac{\partial^m}{\partial \mu^m} P(\{ \nu_i \}, \mu)
$$

\begin{eqnarray}
D_3 =
\sum_j \sum_{l=1}^{\infty} \frac{(-\tau_j)^l}{l!}
(\frac{\partial}{\partial \mu} - 2\mu)^l
L_j  S_{1j} P(\{\nu_i \},  \mu)
-
\nonumber
\\
\nonumber
\sum_j
\frac{\partial}{\partial \nu_j } S_{3j}
 W^+_j  \sum_{m=1}^{\infty} \frac{\tau_j^m}{m!}
\frac{\partial^m}{\partial \mu^m} P(\{ \nu_i \}, \mu)
\end{eqnarray}

\begin{eqnarray}
D_1 =
\\ \nonumber
\sum_{j'}
 W^+_{j'} \alpha_{acc\ j'} \frac{c_{g\ j'}} {2 \sum_j c_j \nu_j}
(\frac{\partial}{\partial \mu}-2\mu)
\frac{\partial}{\partial \mu} P(\{ \nu_i \}, \mu)
+
\\ \nonumber
\sum_{i'}
 W^+_{i'} (1-\alpha_{c\ i'}) \alpha_{acc\ i'} \frac{c_{i'}} {2 \sum_j c_j
\nu_j}
(\frac{\partial}{\partial \mu}-2\mu)
\frac{\partial}{\partial \mu} P(\{ \nu_i \}, \mu)
+
\\ \nonumber
\sum_{i'}
 W^+_{i'} \alpha_{c\ i'} \frac{c_{i'}} {2 \sum_j c_j \nu_j}
(\frac{\partial}{\partial \mu}-2\mu)
\frac{\partial}{\partial \mu}
 P(\{ \nu_i \}, \mu)
-
\\
\nonumber
\sum_j \sum_{l=1}^{\infty} \frac{(-\tau_j)^l}{l!}
(\frac{\partial}{\partial \mu} - 2\mu)^l
 W^+_j  S_{2j}  \frac{\tau_j^l}{l!}
\frac{\partial^l}{\partial \mu^l} P(\{ \nu_i \}, \mu)
\end{eqnarray}

Now we shall estimate the actions of operators $D_1$ - $D_4$.
One can easy note that the differentiation along $\nu_i$  can be estimated
as
$$
\frac{\partial }{\partial \nu_i} P
\sim
\frac{P}{\Delta \nu_i}
$$
which produces the small parameter $1/\Delta \nu_i$. One has to note that
as has been already noted the value $\Delta \nu_i$ differs from the analogous
value in one component case which can lead to the error made in \cite{Djik}.
The values $\Delta \nu_i$ don't estimate the size of the nearcritical
region (The standard definition of the nearcritical region is given by
inequality $|F - F_c| \leq 1$. The infinite tails can be cut off to reduce
the form of the nearcritical region to a rectangular one.). To estimate
the size of the nearcritical region one can take derivatives of the free
energy along $\kappa$, $\xi_i$.
As the result we come to       the following convention: we use  the notation
$1/\Delta \nu_i$ but keep in mind that the real small parameter will be
$1/\Delta \kappa$. Moreover one can not estimate the size of the nearcritical
region by $\Delta \nu_i$ as it was done in \cite{Djik} but has to use
the halfwidths along $\kappa$ and $\xi_i$.

The value of derivative $\partial F / \partial \nu_i$ also contains the
small parameter $1 / \Delta \nu_i$. So the action of $L_i$ can be estimated
as
$$
L_i P \sim  W^+_i \frac{P}{\Delta \nu_i}
$$

The differentiation along $\mu$ doesn't produce any small parameter.
The characteristic value of $\mu$ is $1$.

Now we can calculate the powers of operators $D_1$ - $D_4$. As the result
we see that
\begin{itemize}
\item
Operator $D_4$ is the smallest one. It has the order
$1/(\Delta \nu_i)^2$.
\item
Operator $D_3$ is small. It has the order $1/\Delta \nu_i$.
\item
Operators $D_1$ and $D_2$ have one and the same order $1$. Here operator
$D_1$ has the formal priority because $D_2$ has no terms without factorials
in the denominators.
\end{itemize}

The main problem of the further analysis is that the main operator $D_1
+ D_2$ ensures relaxation to the state which gives zero flow of the embryos
from the precritical to the postcritical region. So, this state leads
to the zero value of the nucleation rate. To overcome this difficulty
one can use the Chapman-Enskog procedure.

\pagebreak

\end{document}